\documentclass[twocolumn,showpacs,preprintnumbers,amsmath,amssymb]{revtex4}
\usepackage{amsmath}
\usepackage{amsfonts}
\usepackage{amssymb}
\usepackage{epsfig}

\begin{document}

\preprint{}
\title{Evolutionary dynamics on rugged fitness landscapes:
exact dynamics and information theoretical aspects}
\author{David B. Saakian$^{1,2}$}
\author{Jos\'e F. Fontanari$^{1}$}
 \affiliation{$^1$Instituto de F\'{\i}sica de S\~ao Carlos,
 Universidade de S\~ao Paulo, Caixa Postal 369, 13560-970 S\~ao Carlos, S\~ao Paulo, Brazil}
 \affiliation{$^2$Yerevan Physics Institute, Alikhanian Brothers St. 2, Yerevan 375036, Armenia }

\date{\today}

\begin{abstract}
The parallel mutation-selection evolutionary dynamics, in which mutation and replication are
independent events, is solved exactly in the case  that  the Malthusian fitnesses associated to the
genomes are described by the Random Energy Model (REM) and by a ferromagnetic  version of the REM.
The solution method uses the mapping of the evolutionary dynamics into a quantum Ising chain
in a transverse field and the Suzuki-Trotter formalism to calculate the transition probabilities
between configurations at different times. We find that in the case of the REM  landscape
the dynamics can exhibit three distinct regimes:  pure diffusion or stasis for short times, depending
on the fitness of the initial configuration,
and a spin-glass regime for large times. The dynamic transition between these dynamical regimes
is marked by discontinuities in the mean-fitness as well as in the overlap with the initial reference
sequence. The  relaxation  to  equilibrium is described by an inverse time decay.
In the ferromagnetic REM, we find in addition to these three regimes, a ferromagnetic regime where the
overlap and the mean-fitness are frozen. In this case, the system relaxes to equilibrium in a finite time.
The  relevance of our results to  information processing aspects of
evolution is discussed.

% A polynomial class (from computation point of view) of evolution
%models is discussed.

\end{abstract}
\maketitle

%%%%%%%%%%%%%%%%%%%%%%%%%%%%%%%%%%%%%%%%%%%%%%%%%%%%%%%%%%%%%%%%%%%%%%%%%%
%
\section{Introduction}\label{sec:Intro}
%
%%%%%%%%%%%%%%%%%%%%%%%%%%%%%%%%%%%%%%%%%%%%%%%%%%%%%%%%%%%%%%%%%%%%%%%%%%

Evolution on complex fitness landscapes has been advanced as a key idea in
 recent mathematical approaches to evolution theory \cite{ka87,Schuster_02}.
Concepts such as  neutral networks and punctuated equilibrium, which are
central to such disparate areas as molecular evolution and paleontology, can
be brought together within that unifying framework. Since frustration
and quenched disorder combined together yield  an almost infallible recipe to generate
complexity \cite{Mezard_87}, the evolutionary dynamics on rugged fitness landscapes
became a research topic in the statistical mechanics of disordered systems.

There are a few reasons to consider spin-glass or random fitness landscapes in the
study of molecular evolution.  Fitness
functions are related to the binding affinity of a molecular replicator (a RNA-like molecule)
to a non-specific replicase \cite{Stein_84}. Given our present incapacity to predict
affinity -- single amino acid replacements may prevent binding altogether, increase
affinity by orders of magnitude, or simply leave it unaffected  \cite{Perelson_95} -- 
the assignment of fitness values chosen at random from some probability distribution 
seems to be the least biased course to introduce fitness in evolution models. In addition, evolution
in any fitness fitness landscape characterized by a finite correlation length will
resemble evolution in a random landscape when viewed at an appropriate coarse-grained
scale of the sequence configuration space \cite{Flyvbjerg_92}. Finally, the
analysis of the evolution on rugged or random fitness landscapes has produced
dynamical patterns, such as punctuated equilibria (see \cite{Sibani_95,Aranson_97}), that
are actually observed in microbial populations \cite{Elena_96}.

Building on a mapping between the infinite-population quasispecies model
\cite{ei71} and an anisotropic two-dimensional  Ising spin model  in which the
time $t$  is one of the lattice dimensions
\cite{Leuthausser_87,Tarazona_92}, the evolutionary version of Derrida's
Random Energy Model (REM) \cite{de80,Gross_84} was solved exactly in the
infinite-time, equilibrium regime \cite{fr93,fr97}. Two distinct phases
were found, corresponding to the selective and non-selective regimes
that characterize a model that exhibits an error threshold transition
(see also \cite{or03}). More recently, the dynamics of this model was
investigated in the limit of strong selection using
 approximate techniques \cite{kr03,kr05,de06,ja07}.
Since strong selection is not the only biologically relevant situation, and
the selective regime is not the only important dynamic regime,  a
more general approach to the dynamics of the evolutionary version of
REM is necessary.

In this contribution, we explore a different mapping between evolutionary dynamics and
statistical physics, namely, the mapping between the parallel mutation-selection
scheme and the quantum Ising chain in a transverse field \cite{Baake_97},
to solve exactly the dynamics on REM-like fitness landscape for all range
of the selection and mutation parameters. This approach was already successfully
used in the analysis of a simpler fitness landscape, the Single-Peak fitness landscape
\cite{sa04} and is based  on the results of Refs.\ \cite{go90,sa93}. A more
recent application was the solution of the evolutionary dynamics  in the
case of symmetric fitness landscapes
(i.e., the fitness is a function of the Hamming distance from a given reference
sequence) \cite{sa08}.

Here we consider two landscapes, the ordinary REM landscape and the ferromagnetic REM
landscape, where one of the energy levels of the ordinary REM is selected and
changed to an energy value lower than the typical ground state energy. 
For both landscapes, we find
that the dynamic behavior for short times depends on
the (Malthusian) fitness of the initial configuration. When the fitness of the initial
configuration is higher than the mutation rate, the dynamics freezes at the initial
configuration, resulting in a pattern
of stasis (provided that the initial fitness is not the global maximum). Otherwise, when the
initial fitness is lower than the mutation rate, the dynamics is characterized by a
regime of pure diffusion in the sequence space.
In the
ordinary REM, we find that these short-time regimes -- diffusion or stasis --
change abruptly to a spin-glass-like regime which is associated
to the equilibrium frozen phase  of the REM. This dynamic transition is signaled by a
discontinuity of the mean fitness as well as of the average overlap with the initial,
reference sequence. Most importantly, we find that these quantities tend
to their equilibrium values as $1/t$ for large time $t$.
In the case of the ferromagnetic REM we find up to four distinct dynamic
regimes: diffusion or stasis, spin glass and ferromagnetic. As before, the transitions between
these regimes are signaled by discontinuities in the biologically relevant observables,
but the system relaxes to  the equilibrium ferromagnetic state in a finite time.

Our results are interesting for the statistical physics aspects  of information
theory as well \cite{so89,b4,b5,b6,b7,sa05}, as REM statistical physics gives a
simple derivation of most of information theory results.
The idea of coding via statistical mechanics is to construct a spin Hamiltonian, which
has a known ground state for a specific choice of the  deterministic spin couplings
(the non-trivial aspect of the problem is that the ground state of the Hamiltonian
should be robust to some degree of noise in those couplings). The
ground state of the Hamiltonian could then be recovered from the starting configuration using
some update dynamics, as in the case of associative memory neural networks \cite{Hopfield_82}.
Thus by solving an evolution model we get as a by-product an analytical decoding dynamics
for optimal codes.

The rest of the paper is organized as follows. In Sect.\ \ref{sec:Chain} we introduce
the evolution equations for the parallel mutation-selection scheme and discuss
its relation with the quantum Ising chain in a transverse field. The basic equations
for the dynamics obtained using the Suzuki-Trotter formalism \cite{Suzuki_76} are introduced also
in that section. In Sect.\ \ref{sec:SP} we review the main results obtained in the analysis
of the Single-Peak landscape \cite{sa04} as they  underlie most of the arguments used in the
solution of the more complex landscapes. In Sects.\  \ref{sec:REM} and \ref{sec:FREM}
we present the exact solution of the evolutionary dynamics of the ordinary REM and of
ferromagnetic REM, respectively. Finally, in Sect.\ \ref{sec:Conc} we summarize our
main results and present some concluding remarks.

%%%%%%%%%%%%%%%%%%%%%%%%%%%%%%%%%%%%%%%%%%%%%%%%%%%%%%%%%%%%%%%%%%%%%%%%%%%%%%

%%%%%%%%%%%%%%%%%%%%%%%%%%%%%%%%%%%%%%%%%%%%%%%%%%%%%%%%%%%%%%%%%%%%%%%%%%
%
\section{Ising quantum chain formulation}\label{sec:Chain}
%
%%%%%%%%%%%%%%%%%%%%%%%%%%%%%%%%%%%%%%%%%%%%%%%%%%%%%%%%%%%%%%%%%%%%%%%%%%

In this contribution we  consider the so-called parallel mutation-selection scheme
in which mutation and selection are considered as independent events \cite{ki_70,Wiehe_95,Hofbauer_88},
i.e.,  mutations can occur at any time during the existence of a sequence, not only
at the moment of replication as assumed in Eigen's molecular quasispecies model \cite{ei71}.
As usual, we represent a molecule or genome of length $N$ by a sequence of binary digits (spins)
$s_k = \pm 1$ with $k=1, \ldots, N$ so  that there are $2^N$ distinct molecules
$S^i \equiv \left ( s^i_1, \ldots, s^i_N \right )$. In the parallel mutation-selection
scheme, the relative frequencies of molecules $i=1,\ldots,2^N$ are given by \cite{ki_70}
\begin{equation}\label{p_i}
\frac{dp_i}{dt} = p_i \left ( r_i - \sum_{j=1}^{2^N} r_j p_j \right ) + \sum_{j=1}^{2^N} m_{ij} p_j
\end{equation}
where $r_{i}$ are Malthusian fitnesses, which can take on positive as well as negative values \cite{Wiehe_95},
and $m_{ij}$ is the mutation rate from  $S^i$ to $S^j$. Since mutations can connect only nearest neighboring sequences
in the $2^N$-dimensional sequence space, we choose  $m_{ij} = \gamma$ if $d \left ( S^i, S^j \right ) = 1$;
$m_{ii} = - N \gamma $ and $m_{ij} = 0$, otherwise. Here $d \left ( S^i, S^j \right )$ is
the Hamming distance between sequences $S^i$ and $S^j$ and $\gamma$ is the mutation rate per site.
As $\sum_i m_{ij} = 0$, the dynamics (\ref{p_i}) maintains the normalization $\sum_i p_i = 1$ for all
$t$. Finally, we recall that  $r_i = f \left ( s_1^i, \ldots, s_N^i \right )$ determines the
so-called fitness landscape.

A key observation at this stage is the finding that the non-linear dynamic system (\ref{p_i}) can be reduced to
a linear system
\begin{equation}\label{x_i}
\frac{dx_i}{dt} = \sum_j H_{ij} x_j
\end{equation}
where $H_{ij} = H_{ji} \equiv r_i \delta_{ij} + m_{ij}$ using the transformation  \cite{Thompson_74}
\begin{equation}
x_i \left ( t \right ) = p_i  \left ( t \right )
\exp \left [ \sum_j r_j \int_0^t d \tau ~ p_j  \left ( \tau \right )  \right ] .
\end{equation}
In practice, we solve the linear system (\ref{x_i}) and then obtain the original sequence
frequencies via the normalization $p_i = x_i/\sum_j x_j$. From the numerical perspective, solution
of this linear system is straightforward, the sole limitation being the exponential increase
in the number of equations with the sequence length $N$.

For certain fitness landscapes, however, the case of infinite length sequences can be
solved analytically thanks to a cunning observation by Baake et al. \cite{Baake_97}, who
realized that the linear system (\ref{x_i}) can be mapped into an Ising quantum chain in a transverse
magnetic field with spin interactions that depend on the specific choice of the fitness
landscape. More pointedly, the linear system (\ref{x_i}) is equivalent to the evolution of the
quantum system described by the Hamiltonian \cite{Baake_97}
\begin{equation} \label{Hamiltonian}
 - \mathcal{H}  = \gamma \left ( \sum_{k=1}^N \sigma^x_k - N \right ) +
f \left ( \sigma^z_1, \ldots, \sigma^z_N \right )
\end{equation}
where $\sigma^{x,z}_k$ stands for the Pauli spin operators acting in site $k$,
i.e., $ \sigma^{x,z}_k = 1 \otimes \ldots 1 \otimes \sigma^{x,z} \otimes 1 \ldots \otimes 1$
with $\sigma^{x,z} $ in the $k$th place.

Introducing
the time evolution operator $ \mathcal{T} \left ( t \right ) = \exp \left ( - \mathcal{H} t \right )$ we can write
a formal expression for the original molecules frequencies, namely,
\begin{equation}
p_j \left ( t \right ) = \frac{1}{\mathcal{N}} \sum_{i=1}^{2^N} Z_{ji} \left ( t \right )  p_i \left ( 0 \right )
\end{equation}
where
$Z_{ji} \left ( t \right )  = \langle S^j | \mathcal{T} \left ( t \right ) | S^i \rangle $
and $\mathcal{N} = \sum_{ji} Z_{ji}  \left ( t \right ) p_i \left ( 0 \right )$ guarantees the correct normalization.
Here $ | S^i \rangle = | \chi_1^i \rangle \otimes \ldots  \otimes  | \chi_N^i \rangle $ where
$ | \chi_k^i \rangle $ is an eigenstate of $\sigma^{z}_k$.

Since there is an equivalence  between the quantum  Ising  chain, such as that
described by the Hamiltonian (\ref{Hamiltonian}), and a  classical anisotropic two-dimensional
Ising spin model \cite{Suzuki_76}, then Baake et al.'s observation shows that
there   is a mapping between the parallel mutation-selection  evolution scheme
and the two-dimensional Ising model. It is interesting that a similar result
holds for   Eigen's quasispecies model as well  \cite{Leuthausser_87,Tarazona_92}.

The challenge here is to calculate $Z_{ji} \left ( t \right ) $ which is proportional to the probability
that state $ | S^i \rangle$  transitions to state $| S^j \rangle$ in a time interval of length $t$.
Henceforth we will refer to $Z_{ji}$ as the transition amplitude between those states.
In the case that the spins interaction, i.e., the term  $f \left ( \sigma^z_1, \ldots, \sigma^z_N \right )$
in Eq.\ (\ref{Hamiltonian}),  can be neglected we have
$\mathcal{T} \to \mathcal{T}_{diff} = \exp \left [ \gamma \sum_i \left ( \sigma^x_i-1 \right ) \right ]$
so that $Z_{ji}$  can  be readily evaluated \cite{Suzuki_76} (see also \cite{sa04}),
\begin{equation}\label{Tdif}
\langle S^j | \mathcal{T}_{diff} \left ( t \right ) | S^i \rangle  =
\exp \left [ N \left ( \phi \left ( m, t \right ) - \gamma t \right ) \right ]
\end{equation}
where
\begin{equation}\label{phi}
\phi \left ( m,t \right )=\frac{1+m}{2} \ln \cosh \left ( \gamma t \right )
+ \frac{1-m}{2} \ln \sinh \left ( \gamma t \right )
\end{equation}
and $m$ is the overlap between configurations $S^i$ and $S^j$, i.e.,
$m = \sum_k s_k^i s_k^j /N$. On the other hand, in the case the interactions are dominant
we have
$\mathcal{T} \to \mathcal{T}_{int} = \exp \left [  - N \gamma t  +
f \left ( \sigma^z_1, \ldots, \sigma^z_N \right ) t \right ]$ and so
\begin{equation}\label{Tint}
\langle S^j | \mathcal{T}_{int} \left ( t \right ) | S^i \rangle  =
\exp \left [ - N  \gamma t + f \left ( S^i \right ) t \right ] \delta_{ij} .
\end{equation}
For a given ensemble of initial configuration $S^i$ our aim  is to determine which configurations
$S^j$ maximize the transition amplitude $Z_{ji} \left ( t \right )$.
From Eq.\ (\ref{Tdif}) we can already realize that the answer will depend only on the overlap
between these two configurations and so the problem is reduced to finding the overlap
$m$ that maximizes $Z_{ij}$. Henceforth we will refer to this maximum as
$Z = \max_j Z_{ji} $, thus omitting, for the sake of simplicity, the dependence on the initial configuration index $i$.
Of course, because of the large $N$ limit we  have $Z = \sum_j Z_{ji} $  as well.

The simplest fitness landscape for which the  transition amplitudes
$Z_{ji}$ can be calculated exactly is the so-called Single-Peak (SP)
fitness landscape. (The SP happens to be also the most studied fitness
landscape in the quasispecies literature.) In this case, there is a single configuration
-- the master sequence  $S^0$ --
with a high fitness value  $N J_0$, whereas all other configurations have their fitness values
set to zero. By choosing the master sequence as $S^0 = \left ( 1,\ldots, 1 \right )$, the spin
interactions for the SP landscape can be written as  \cite{sa04}
\begin{equation}\label{f_SP}
f_{SP} \left ( \sigma^z_1, \ldots, \sigma^z_N \right ) = N J_0 \left ( \sum_k \sigma_k^z / N \right )^p
\end{equation}
in the limit $p \to \infty$. What makes the SP problem analytically solvable was the
remarkable  finding that $Z$  can be written in a factorized form
\begin{equation}\label{dec}
 Z = \sum_j \int_0^t  d t_1 \langle S^j | \mathcal{T} _{int} \left ( t-t_1 \right ) | S^j \rangle
\langle S^j | \mathcal{T}_{diff} \left ( t_1  \right ) | S ^i \rangle .
\end{equation}
We refer the reader to the Appendix of Ref.\ \cite{sa04} for the detailed derivation of this result,
which is based on the Trotter-Suzuki scheme \cite{Suzuki_76} (see also \cite{go90,sa93}) that
introduce infinitely many intermediate time steps between the initial and the final configurations.
The key point for the factorization is the large $p$-spin interaction [see Eq.\ (\ref{f_SP})]
in the Hamiltonian, which results in a ground-state configuration with fitness much greater than
the fitness of typical configurations. This is obvious for the SP landscape (except for the master,
all configurations have zero fitness), but also holds for the REM landscape for which typical configurations
have fitness on the order of $N^{1/2}$ whereas the ground-state has fitness on order of $N$.

%%%%%%%%%%%%%%%%%%%%%%%%%%%%%%%%%%%%%%%%%%%%%%%%%%%%%%%%%%%%%%%%%%%%%%%%%%
%
\section{SP fitness landscape}\label{sec:SP}
%
%%%%%%%%%%%%%%%%%%%%%%%%%%%%%%%%%%%%%%%%%%%%%%%%%%%%%%%%%%%%%%%%%%%%%%%%%%

It is instructive to present the results for $Z$ in the case of the single-peak landscape
defined in the previous section and  investigated in Ref.\ \cite{sa04}. In particular, here we
focus on the time dependence of the average overlap $m$ between the configurations
at time $t$ and the initial configuration, a quantity which was not explored
in that seminal work.
First, we replace
the sum over the final configurations $S^j$ by an integral over all possible values of
the overlap between the final and the initial configurations, taking into account that for large $N$
there are
$\rho \left ( m \right ) = \exp \left [ N h \left ( m \right ) \right ] $ distinct configurations
for a fixed overlap $m$,
where
\begin{equation}\label{h}
 h \left ( m \right ) =  -\frac{1+m}{2} \ln \frac{1+m}{2} - \frac{1-m}{2} \ln \frac{1-m}{2} .
\end{equation}
Then we  use Eqs.\ (\ref{Tdif}) and  (\ref{Tint}) to rewrite $Z$ as
\begin{equation}\label{Z_SP}
 Z = \int_{-1}^1 dm  \int_0^t  d t_1
\exp \left [  N F_{SP} \left ( m, t_1 \right ) \right ]
\end{equation}
where
\begin{equation}\label{F_SP}
 F_{SP } = h \left ( m \right )  + \phi \left ( m,t_1 \right ) -\gamma t + J_0 \left ( t-t_1 \right )  .
\end{equation}

The integrals over $m$ and $t_1$ can be easily carried out for large $N$ using Laplace's method as
only the contribution of the maximum of $F_{SP}$ is relevant for the evaluation of $Z$.
We find a maximum at the extreme of the $t_1$ integration interval, i.e., $t_1 = t$, which, according to
Eq.\ (\ref{dec}), corresponds to a regime of pure diffusion in the sequence space. In fact, maximization
of $F_{SP}$ with respect to $m$ for $t_1=t$ yields
\begin{equation}\label{m_D}
m = \exp \left ( -2\gamma t \right ) ,
\end{equation}
from where we obtain $F_{SP} = 0$.
Next, we consider the maximum within the integration intervals.
The condition of maximum  with respect to $t_1$ yields
\begin{equation}\label{qua_SP}
\left ( 1+m \right ) \tanh \left ( \gamma t_1 \right ) + \frac{ 1-m }{\tanh \left ( \gamma t_1 \right )}
 -\frac{2J_0}{\gamma}   = 0 .
\end{equation}
This is a quadratic equation for the unknown $\tanh \left ( \gamma t_1 \right )$ which has real
solutions provided that $J_0  \geq \gamma$ regardless of the value of the other unknown, $m$.
This is an evidence that this solution describes the selective regime of the parallel selection-mutation
evolution model. Now, maximization of $F_{SP}$ with respect to $m$ yields $m = \exp \left ( -2 \gamma t_1 \right )$.
The problem is that inserting this expression into Eq.\ (\ref{qua_SP}) results in $J_0 = \gamma$, and so for
this particular situation $m$ can take on any value, whereas $t_1$ is given by Eq.\ (\ref{qua_SP})
and $F_{SP}=0$.

To understand what is happening here we recall that the average overlap between the sequences in the
quasispecies distribution at equilibrium and the master sequence equals  exactly $1$ in the $N \to \infty$ limit.
This is so despite the fact that the master sequences comprise only the fraction $ 1- \gamma /J_0$
of the total number of sequences at equilibrium. In the same  vein, the overlap $m$ between the initial
configuration and the equilibrium configurations become identical to the overlap
$m_0$ between the initial configuration and the master sequence, which is an
arbitrary quantity defined by the initial conditions of the system. This explains the fact that
the overlap $m$ in Eq.\ (\ref{qua_SP})  is not specified by the maximization conditions for $J_0 = \gamma$:
it is determined by the initial conditions, $m = m_0$.

%
%------------------------------------------------------------------------
\begin{figure}
\centerline{\epsfig{width=0.52\textwidth,file=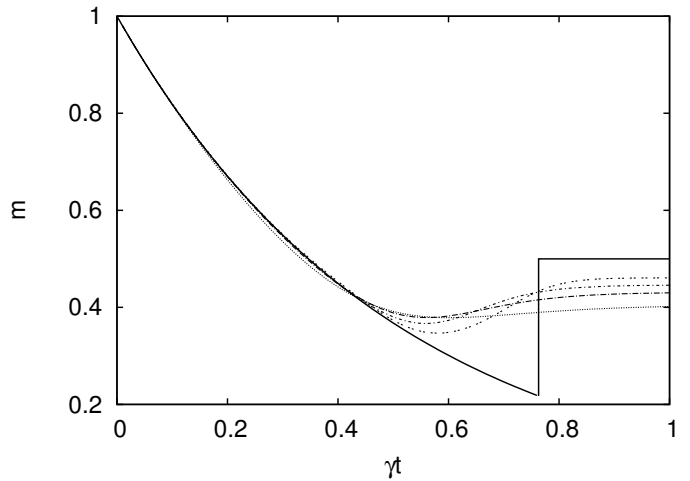}}
\par
\caption{Average overlap with the initial configuration as function of the scaled time $\gamma t$
for $J_0/\gamma=2$, $m_0 = 0.5$ and (broken lines from bottom to top  at
$\gamma t_{df} \approx 0.763 $) $N=8,  12, 16$ and $24$.
The thick solid line is the theoretical prediction.}
\label{SP:1}
\end{figure}
%------------------------------------------------------------------------
%

To take into account the possibility that the dynamics reaches the close neighborhood of the
master sequence, and so
the overlap $m \left ( t \right )$ freezes at the value $m_0$,   in a finite time we now
calculate the transition amplitude  $Z_{0i}$ between $S^i$ and the master sequence $S^0$
in  the time interval $t$, given that $m_0 = \sum_k s_k^i s_k^0/N$. Since the final state is
fixed the entropic term (\ref{h}) must be dropped and we find
\begin{equation}\label{Z_0i}
Z_{0i}  \propto \exp \left [ N \left ( \phi \left ( m_0,t_1 \right ) -\gamma t + J_0 \left ( t-t_1 \right )
\right ) \right ]
\end{equation}
where $t_1$ is given by Eq.\ (\ref{qua_SP}) with $m=m_0$. Recalling that $F_{SP}=0$
for the diffusive regime (\ref{m_D}), the selective regime takes over at time $t$
such that $\phi \left ( m_0,t_1 \right ) -\gamma t + J_0 \left ( t-t_1 \right ) > 0$ \cite{sa04}.
Figure \ref{SP:1}, which exhibits the time-dependence of the overlap $m$, summarizes these results for
a particular choice of $J_0/\gamma$ and $m_0$. The jump of the overlap $m$ at $\gamma t_{df}$
signals the transition between the diffusive and the selective (frozen) dynamic regimes.
We note that for the value of $J_0/\gamma = 2$ used in the figure, the overlap jumps up if $m_0 > 0.112$
and jumps down otherwise. The transition between the two regimes is continuous for $m_0 \approx 0.122$.
As illustrated in Fig.\ \ref{SP:1}, the results of the numerical integration of the system of equations (\ref{p_i})
using the Runge-Kutta method show a clear trend to  converge to
the theoretical predictions as the sequence length $N$ increases.

To conclude this brief overview of the parallel mutation-selection dynamics for the SP landscape
we mention that the mean fitness $R$ of the population is zero in the diffusive regime and
$R = J_0 - \gamma$ (i.e., $J_0$ times the frequency of  master sequences in the population,
$ 1 - \gamma/J_0$) in the selective phase.

%%%%%%%%%%%%%%%%%%%%%%%%%%%%%%%%%%%%%%%%%%%%%%%%%%%%%%%%%%%%%%%%%%%%%%%%%%
%
\section{REM fitness landscape}\label{sec:REM}
%
%%%%%%%%%%%%%%%%%%%%%%%%%%%%%%%%%%%%%%%%%%%%%%%%%%%%%%%%%%%%%%%%%%%%%%%%%%

In this case the fitness landscape is given by \cite{de80,Gross_84}
\begin{equation}\label{p_coup}
f \left ( s_1, \ldots, s_N \right ) =
\sum_{i_1 < i_2 \ldots < i_p} J_{i_1 \ldots i_p} s_{i_1} \ldots s_{i_p}
\end{equation}
where the couplings $J_{i_1 \ldots i_p}$  are Gaussian distributed random variables
of zero mean and variance $\langle J^2_{i_1 \ldots i_p} \rangle =
J^2 p!/\left ( 2 N^{p-1} \right )$. Taking the limit $p \to \infty$ in Eq.\
(\ref{p_coup}) results in
$2^N$ independent energy levels, $E \equiv -  f \left ( s_1, \ldots, s_N \right )$,
distributed by a Gaussian distribution
\begin{equation}\label{Gaussian}
w \left ( E \right ) =  \frac{1}{J\sqrt{\pi N }} \exp \left ( -\frac{E^2}{NJ^2} \right ).
\end{equation}
The equilibrium statistical mechanics of the quantum Ising model in a transverse
field, Eq.\ (\ref{Hamiltonian}),  with spin interactions given by Eq.\ (\ref{p_coup})
was studied in Ref.\ \cite{go90}. In the zero-temperature limit, which is the limit relevant
to our analysis, there are two phases, namely,
a spin-glass or frozen phase that occurs for large $J$, and a paramagnetic phase. The discontinuous
transition between these phases takes place at  $ J/\gamma = 1/\sqrt{\ln 2} \approx  1.201 $ \cite{go90}.
Clearly, within our evolutionary interpretation of the model, this phase transition corresponds
to the error threshold phenomenon: for mutation rates greater than $J \sqrt{\ln 2}$ the adaptive
information stored in the fitness landscape no longer affects the frequencies of sequences in the
population, which become uniformly distributed.

Because of the presence of the quenched random variables $J_{i_1 \ldots i_p}$, the
derivation of the equations for the dynamics of the REM is more involved compared with that
for the SP landscape. A rigorous derivation of the factorization (\ref{dec})
can be done using the Trotter-Suzuki scheme \cite{Suzuki_76,go90,sa93} as in the SP fitness
landscape \cite{sa04}. Here we give a qualitative derivation based on the general
results described in Sect.\ \ref{sec:Chain}.

Effectively, we simply must  replace $f \left ( S^i \right )$ in Eq.\ (\ref{Tint}) by $E_i$ (i.e.,
by the energy of the initial configuration $S^i$), which  amounts to replacing $J_0 $ in Eq.\ (\ref{F_SP})
by $E_i/N$. Since our final  results must be averaged over the energies of the final configurations $S^j$
[see Eq.\ (\ref{dec})] we have
\begin{eqnarray}\label{Z1}
 Z & = & \int_0^t d t_1 \int _{-1}^1 dm \int \frac{dE}{J \sqrt{\pi N}}~
\theta \left [ h \left (m \right )- \left ( \frac{E}{NJ} \right )^2 \right ]
\nonumber \\
&  &
 \times  \exp  \left [ - \frac{E^2}{N J^2}   + E \left ( t - t_1 \right )
\right ] \nonumber \\
&  &
\times \exp  \left [ N h \left ( m \right ) + N \phi \left ( m,t_1 \right )  - N \gamma t \right ] .
\end{eqnarray}
Here the theta function enforces the constraint
\begin{equation}  \label{const}
h \left ( m \right )- \left ( \frac{E}{NJ} \right )^2 \geq 0
\end{equation}
which guarantees that the average number of configurations with energy
$E$ and overlap $m$ with the initial configuration, given by
$\exp \left ( N h(m) - E^2/N J^2 \right )$,
is exponentially large, so that $Z$
becomes a self-averaging quantity \cite{de80,fr97}. Hence
Eq.\ (\ref{Z1}) describes a typical situation of the dynamics at a fixed time $t$. Strictly,
this equation is valid when the  energy of the initial configuration $E_{i}$ is larger
then $-\gamma$. We will discuss later the corrections necessary to describe the
case where this condition is violated. In addition, we assume
that the overlap $m_0$ between the ground-state configuration and the initial configuration
is zero.

The crucial step now is to evaluate the integral over $E$ in Eq.\ (\ref{Z1}) via Laplace's
integration for fixed $m$ and $t_1$. Noting that the result of the integration depends
on whether the critical point $E^* = NJ^2 \left ( t-t_1 \right )/2$ satisfies or not the constraint (\ref{const})
and dropping trivial multiplicative  factors, we rewrite $Z$ as
\begin{equation}\label{Z_D}
Z_D = \int_0^t d t_1 \int _{-1}^1 dm ~ \exp \left [ N F_{D} \left ( m,t_1 \right ) \right ]
\end{equation}
where
\begin{equation}\label{F1}
F_{D}= h \left ( m \right )+ \phi \left ( m,t_1 \right )-\gamma t +
J^2 \left ( t-t_1 \right )^2/4
\end{equation}
provided that
\begin{equation}\label{V_D}
h \left ( m \right )- J^2 (t-t_1)^2/4 > 0 ,
\end{equation}
or
\begin{equation}\label{Z_SG}
Z_{SG} = \int_0^t d t_1 \int _{-1}^1 dm ~ \exp \left [ N F_{SG} \left ( m,t_1 \right ) \right ]
\end{equation}
where
\begin{equation}\label{F2}
 F_{SG}= \phi \left ( m,t_1 \right )- \gamma t + J \sqrt{h(m)} \left ( t-t_1 \right )
\end{equation}
in the case that
\begin{equation}\label{V_SG}
h \left ( m \right )- J^2 (t-t_1)^2/4 < 0 .
\end{equation}
Equation  (\ref{Z_SG})
results when the critical point $E^*$ is outside the energy integration interval,
and so the argument of the exponential is maximized by  the energy $E$ at
the extreme of that interval, namely, $E = NJ \sqrt{h \left ( m \right )}$. As
before, for large $N$ we need to find the values of $m$ and $t_1$ that maximize
the  arguments of the exponentials, Eqs. (\ref{F1}) and (\ref{F2}).
The correct solution is then the one that corresponds to the largest of $Z_D$
and $Z_{SG}$, i.e., $Z = \max \left \{ Z_D,Z_{SG} \right \}$.

We begin  with the analysis of $Z_D$, Eq.\ (\ref{Z_D}). Calculation of the extreme of
$F_{D} \left(m,t_1 \right )$ with respect
to $m$ and $t_1$ yields $m = \exp \left ( -2 \gamma t + 4 \gamma^2/J^2 \right )$ and
$t_1 = t - 2\gamma/J^2$, so that $F_{D} = -\gamma^2/J^2 < 0$. Next we must evaluate $F_{D}$
at the upper extreme of the $t_1$ integration interval, i.e., at $t_1 = t$. This is
the pure diffusion regime discussed in Sect.\ \ref{sec:SP} which results in
Eq.\ (\ref{m_D}) and  $F_{D} = 0$.
The lower extreme $t_1 = 0$ yields $F_{D} \to -\infty$ and
the extremes of the $m$ integration interval (i.e., $m = \pm 1$) need not be considered
because $h \left ( m = \pm 1 \right ) = 0 $ and so the condition (\ref{V_D}) is violated.
We note that the solution given by Eq.\ (\ref{m_D}), which describes the pure drift
or diffusion in the sequence space, exists for all parameter values since
$h \left ( m \right ) \geq 0$ and so
the inequality (\ref{V_D}) is always satisfied. In addition, since solution (\ref{m_D}) yields the
largest value of the exponent $F_{D}$, the other
solutions must be discarded.

We turn now to the analysis of $Z_{SG}$, Eq.\ (\ref{Z_SG}). As before, we start by the maximization of
$F_{SG} \left(m,t_1 \right )$ with respect to both integration variables, $m$ and $t_1$. At the maximum,
we find that the values of these variables are given by the solution of the equations
\begin{equation}\label{m_int}
\ln \tanh \left ( \gamma t_1 \right ) + \frac{J}{2} \frac{\left ( t-t_1 \right )}
{ \sqrt{ h \left ( m \right ) }} \ln \left ( \frac{1+m}{1-m} \right )=0
\end{equation}
and
\begin{equation}\label{quadrac}
\left ( 1+m \right ) \tanh \left ( \gamma t_1 \right ) + \frac{ 1-m }{\tanh \left ( \gamma t_1 \right )}
 -\frac{2J}{\gamma}  \sqrt{h \left ( m \right )} = 0.
\end{equation}
These equations have to be solved numerically, but the solution is simple because
Eq.\ (\ref{quadrac}) can be rewritten as a quadratic  equation  $y \equiv \tanh \left ( \gamma t_1 \right ) < 1$
which then can be written explicitly in terms of the unknown $m$.  Regardless of the value of $m$, this
quadratic  equation has real solutions
provided that $J/\gamma > 1/\sqrt{\ln 2}$ and so we identify this dynamic regime with the
frozen spin-glass phase of the quantum  version of REM \cite{go90}. In addition,  in the case that both roots of
$y$ are physical (i.e., less than 1), our numerical analysis
indicates that we should always choose the smaller root since it corresponds
to the largest value of the exponent $F_{SG}$.

To conclude the analysis of $Z_{SG}$, we must consider the contributions
from  the extremes of the integration intervals. The extreme $t_1 = t$ is discarded
because it violates condition (\ref{V_SG}),  whereas the contribution of $t_1 = 0$ can
be ignored because it yields $F_{SG} \to -\infty$. Regarding the extremes $m = \pm 1$, we
find that in this case $F_{SG}$ is maximum when $t_1$ takes on its extreme value, $t_1 =t$.
This  corresponds to the contribution from the border
$h \left ( m \right )- J^2 (t-t_1)^2/4  = 0 $ which we will discuss in detail in the
Appendix \ref{App_B}.  Our numerical analysis indicates, however, that
the border contribution  can be ignored  since it  yields an exponent $F_{B}$
[see Eq.\  (\ref{F3})] which is
always  smaller than the exponents obtained using the solution of the Eqs.\ (\ref{m_int}) and
(\ref{quadrac}).

In addition to the average overlap $m$ between the initial and the
configuration at time $t$, we can calculate the time dependence of the mean fitness
$R$ of the sequence population as well. The reasoning to derive $R$ is sketched as follows.
For the diffusive regime we have $R=0$ since  the average fitness of any large
sample of configurations visited in this regime is clearly zero for the REM fitness
landscape. To estimate the mean fitness in the selective regime we just  note the
equivalence between the results for the single-peak landscape [see Eqs. (\ref{Z_SP}) and
(\ref{F_SP})] and for the selective phase [see Eqs. (\ref{Z_SG}) and
(\ref{F2})] if we  identify $J_0$ with an effective, time-dependent single-peak fitness  value
$J_{eff} = J \sqrt{h \left ( m \right )}$. (Note that for $t \to \infty$ we have $m \to 0$ so that
$J_{eff} \to J \sqrt{\ln 2} $, which is the ground-state fitness value of the REM.) Since the
population is formed by  master copies with fitness value $N J_{eff}$  as well as by  clouds of mutants
with much smaller fitness (on the  order of $N^{1/2}$) the mean fitness of the population in the
selective regime becomes
\begin{equation}\label{R_REM}
 R = J \sqrt{h \left ( m \right )}- \gamma ,
\end{equation}
in accord with  the well-known result for the (parallel) version of the single-peak landscape.

In summary, for fixed $J/\gamma$ and $\gamma t$ we must solve the saddle-point equations
(\ref{m_int}) and  (\ref{quadrac}) to obtain the exponent $F_{SG}$ (as well the saddle-point
equation (\ref{border}) given in Appendix \ref{App_B}, but we have already mentioned that its
contribution  must be discarded) and then compare with the exponent of the diffusive regime $F_{D} = 0$.
If $F_{SG} > 0$ we pick the value of $m$ associated to the selective regime, otherwise we pick the
diffusion solution given by Eq.\ (\ref{m_D}).

\subsection{Analysis of the results}

%------------------------------------------------------------------------
\begin{figure}
\centerline{\epsfig{width=0.52\textwidth,file=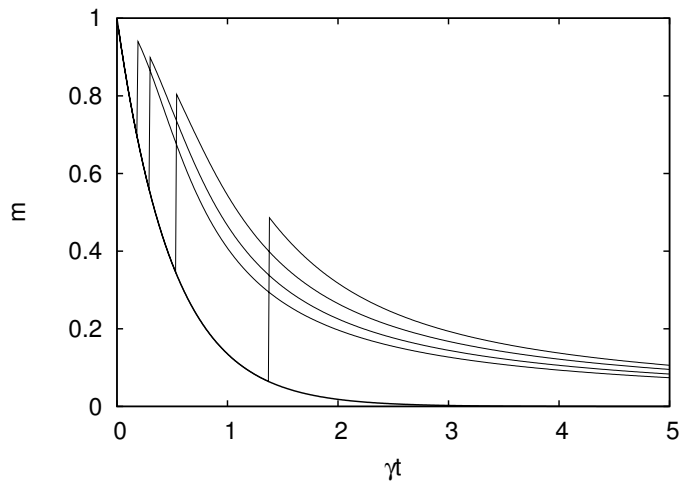}}
\par
\caption{Average overlap with the initial configuration as function of the scaled time $\gamma t$
for (thin solid lines from top to bottom at $\gamma t = 2$) $J/\gamma = 2, 3, 4$ and $5$.
The  thick solid line is the function $\exp \left ( - 2 \gamma t \right )$ which describes
the overlap in the diffusive regime, $J/\gamma \leq  1/\sqrt{\ln 2}$.
}
\label{REM:1}
\end{figure}
%------------------------------------------------------------------------
%

Figure \ref{REM:1} illustrates the typical time evolution of the average overlap $m$ with the initial configuration.
We recall that the fitness of this initial configuration must be less than $\gamma$ and its overlap with
the ground-state configuration must be zero. As expected, for small $\gamma t$ the diffusive regime
dominates and so $m$ is given by Eq.\ (\ref{m_D}). As $\gamma t$ increases further, the selective
regime takes over rather abruptly, as shown by the discontinuity of the overlap $m$ at a critical
time value $\gamma t_{ds}$. This bizarre behavior, which  occurs also in the single-peak fitness landscape,
is consequence of our characterization of the dynamics in a very large-dimensional sequence space by
a single parameter: no  such discontinuous behavior is observed when following the time
evolution of the individual sequence frequencies, $p_i$ for  $i=1,\ldots,2^N$.

\begin{figure}
\centerline{\epsfig{width=0.52\textwidth,file=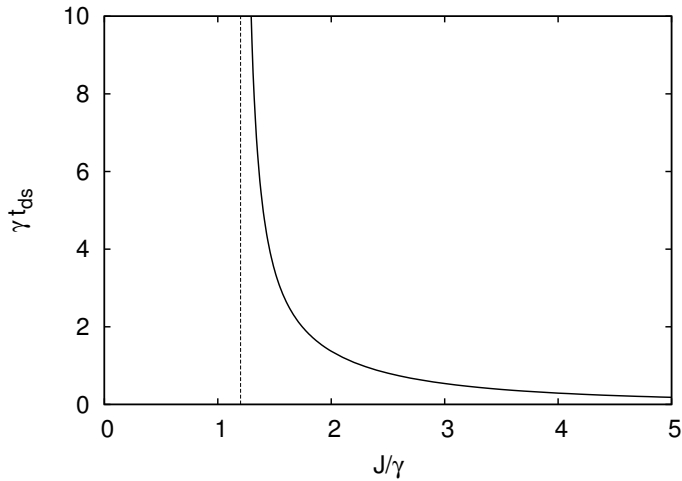}}
\par
\caption{Scaled critical time $\gamma t_{ds}$ at which the discontinuous dynamical transition between the diffusive and
the selective regimes takes place as function of the dimensionless parameter $J/\gamma$. For
$J/\gamma \leq  1/\sqrt{\ln 2}$ only the diffusive regime occurs. The selective regime is dominant
in the region $t > t_{ds}$ (i.e., above the solid line).
}
\label{REM:2}
\end{figure}

\begin{figure}
\centerline{\epsfig{width=0.52\textwidth,file=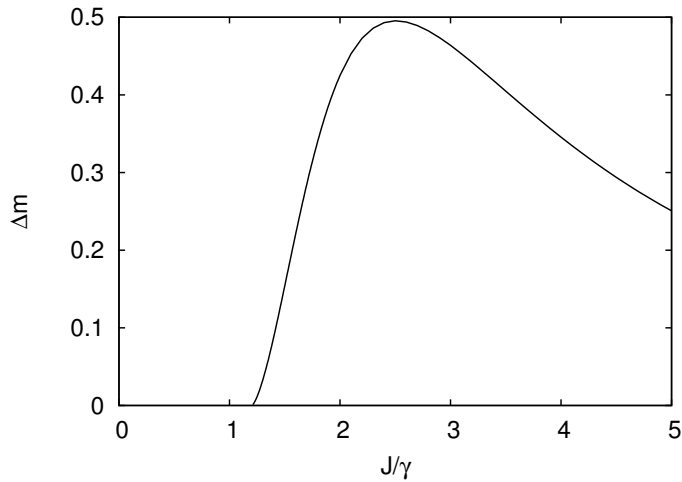}}
\par
\caption{Size of the overlap discontinuity $\Delta m$ at $t=t_{ds}$ as function of the dimensionless parameter $J/\gamma$.
The maximum of  this curve occurs at  $J/\gamma \approx 2.504$. For
for large $J/\gamma$ we find that $\Delta m $ vanishes as $\left ( J/\gamma \right )^{-2}$.}
\label{REM:3}
\end{figure}

Since the properly scaled critical time $\gamma t_{ds}$
at which  the discontinuity of the overlap $m$ (and, consequently, of the mean fitness $R$) takes place
can be used to separate
the regions of validity of the two distinct dynamic regimes,  in Fig. \ref{REM:2} we present the
dynamic `phase diagram' of the parallel evolutionary version of REM. As expected, $\gamma t_{ds}$ diverges
as the $J/\gamma$ approaches the value $1/\sqrt{\ln 2}$ which yields the equilibrium phase
boundary between the paramagnetic and the frozen spin glass phases. For large $J/\gamma$
we find that $\gamma t_{ds}$   vanishes as $\left ( J/\gamma \right )^{-2}$. Also of interest
is the size of the overlap jump at $\gamma t_{ds}$, shown in Fig.\ \ref{REM:3}. The fact that this
quantity exhibits a maximum could already be inferred from Fig.\ \ref{REM:1}, since
the overlap $m$ tends to $1$ or $0$ in both dynamic regimes when $J/\gamma$ approaches its
extreme values.

%
%------------------------------------------------------------------------
%------------------------------------------------------------------------
%

%
%------------------------------------------------------------------------
%------------------------------------------------------------------------
%

\subsection{Numerical integration}

\begin{figure}
\centerline{\epsfig{width=0.52\textwidth,file=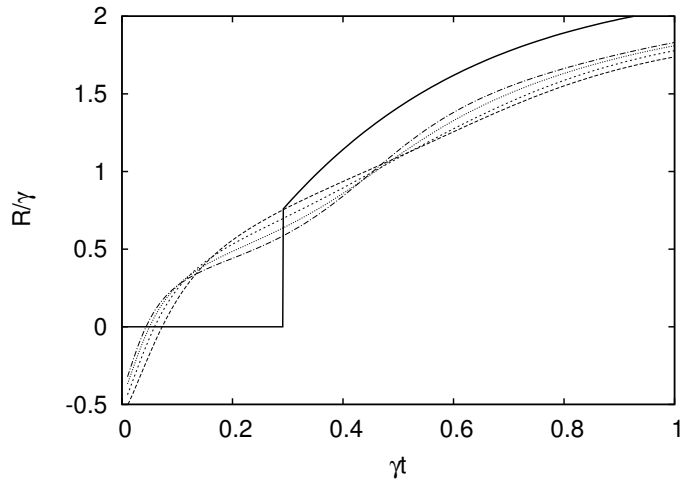}}
\par
\caption{Scaled mean fitness $R/\gamma$ of the REM landscape as function of the scaled time $\gamma t$
for $J/\gamma=4$ and (broken lines from top to bottom at $\gamma t_{ds} \approx 0.292$) $N=12,  16, 20$ and $24$.
The thick solid line is the theoretical prediction.}
\label{REM:4}
\end{figure}

\begin{figure}
\centerline{\epsfig{width=0.52\textwidth,file=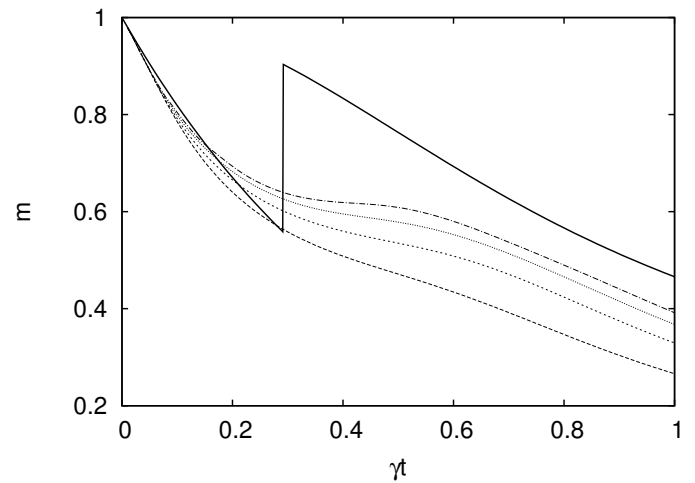}}
\par
\caption{Average overlap with the initial configuration as function of the scaled time $\gamma t$
for $J/\gamma=4$ and (broken lines from bottom to top  at $\gamma t_{ds} \approx 0.292$) $N=12,  16, 20$ and $24$.
The thick solid line is the theoretical prediction.}
\label{REM:5}
\end{figure}

To complement our theoretical analysis, which is exact for infinite sequence lengths,
we have carried the direct numerical integration of the linear system of ordinary
equations (\ref{x_i}) for sequence lengths
up to $N=24$ using the fourth-order Runge-Kutta integrator \cite{Recipes}. The stability of
the numerical procedure  benefited greatly from the fact the differential equations are linear.
For $N \leq 10$ we can find all eigenvectors
and eigenvalues of  the symmetric matrix $H$ and so solve the dynamics
exactly for any $t$ within an arbitrarily high numerical precision. Of course, the
two numerical methods yield identical results provided that $\gamma t$ is not too
large.  Figures \ref{REM:4} and \ref{REM:5} summarize our numerical results for the
$J/\gamma = 4$. For each time $t$ the data in these figures represent the
average over $10^4$ independent samples. The samples differ by the fitness values
assigned to each configuration. For all samples the initial configuration was such as
to have fitness value less than $\gamma$ and zero overlap with the ground-state configuration.

%
%------------------------------------------------------------------------
%------------------------------------------------------------------------
%
%
%------------------------------------------------------------------------
%------------------------------------------------------------------------
%

\begin{figure}
\centerline{\epsfig{width=0.52\textwidth,file=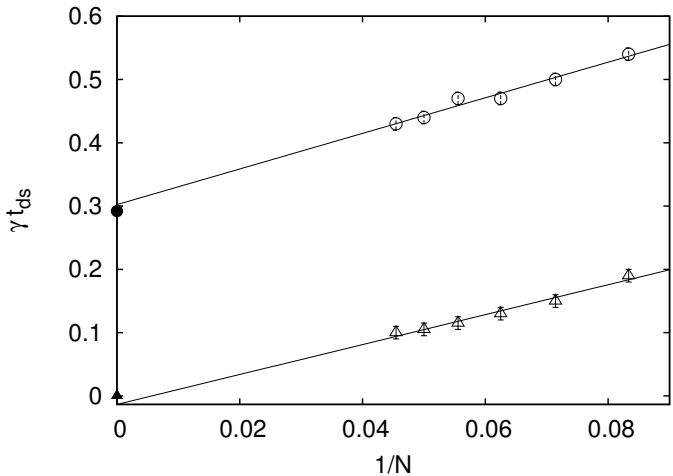}}
\par
\caption{Values of $\gamma t$ at which the mean fitnesses of
sequences of length $N$  and $N+2$  intersect
shown as function of $1/N$ for $N=12,14,16,18,20$ and $22$ for $J/\gamma=4$. The symbols  $\triangle$ and $\bigcirc$ identify
the first and the second crossings, respectively (see Fig.\ \ref{REM:4}), whereas the filled symbols indicate
the theoretical predictions. The solid lines are the linear fittings used to obtain the extrapolated
values at $1/N = 0$.}
\label{REM:6}
\end{figure}

Figure \ref{REM:4} is reassuring because the crossings of the lines for distinct $N$ indicate
the onset of a threshold phenomenon in the thermodynamic limit $N \to \infty$. In particular,
to reproduce the analytical predictions, the first intersection point should tend to $\gamma t = 0$
whereas the second should tend to $\gamma t_{ds} \approx 0.292$. To verify whether our  data exhibit
the correct trend, we show in Fig.\ \ref{REM:6} the values of $\gamma t$ at which the mean fitness curves
intersect for successive values of $N$. The extrapolation to $N \to \infty$ yields $\gamma t = -0.01 \pm 0.01$
for the first crossing and  $\gamma t_{ds}  = 0.30 \pm 0.01$  for the second crossing. The agreement with the
theoretical prediction is excellent, given the  short sequence lengths used in the numerical integration.
Oddly enough, the dependence of the overlap $m$  on the sequence length $N$, shown in Fig.\ \ref{REM:5}, does not
exhibit the characteristic crossings that signalize the onset of a threshold phenomenon in the thermodynamic limit,
although  the curve for $N=24$ already begins to take a shape that resembles the theoretical prediction. It seems
that much larger sequence lengths are needed in order we can obtain clear
evidence of a threshold phenomenon using the overlap data. We refer the reader to Ref.\ \cite{Campos_98} for a
full analysis of the finite size effects of the error threshold transition of the quasispecies model.

%------------------------------------------------------------------------
%------------------------------------------------------------------------
%

\subsection{High-fitness initial configuration}

To complete our analysis of the calculation of $Z$   we consider now
the situation in which the energy of the initial configuration $E_i$
is such that $E_i  < -\gamma$, i.e., this configuration has a
relatively high fitness. In this case, we find a new  paramagnetic
(but non-diffusive) regime where the  system stays in the original
configuration with a probability proportional to [see Eq.\
(\ref{Tint})]
\begin{equation}\label{initial}
  Z = \exp \left [ N \left ( - E_i - \gamma \right ) t \right ]
\end{equation}
and mean fitness $R = -E_i$. Since the argument of the exponential is always positive,
this new regime replaces the diffusive regime altogether. As $t$ increases, it eventually
becomes replaced by the selective regime at some threshold time $t_{ds}' > t_{ds}$. However, the dependence of
$t_{ds}'$ on the particular value $E_i$ makes this case rather unattractive, as compared with
the case where the initial configuration has a low-fitness value.

Finally, we note that the probability that $E_i  < -\gamma$
is $\frac{1}{2} \mbox{erfc} \left ( \gamma/J \sqrt{N} \right )$ which tends to $1/2$ for
large $N$, hence we could more simply distinguish the two
situations -- high and low initial fitness -- by verifying whether the energy of the
initial configuration is positive or negative. More importantly, these two cases are equally likely and
our penchant for the low-fitness initial configuration here is justified only by
the generality of the results obtained in that case.

\subsection{Relaxation to equilibrium}

The approach to the equilibrium state as  $\gamma t \to \infty$ is particularly
interesting because it is related to the speed of evolution, i.e., how
long it takes for a random sequence to reach the global maximum of
a rugged fitness landscape. In contrast with the finite-time dynamics described before,
the results of the asymptotic analysis do not depend
on the specific value of the fitness of the initial configuration. We  still require, however, that
the initial configuration has zero overlap with the ground state.

As we focus on the limit of large $\gamma t$, the relevant equations
to describe the system dynamics are Eqs.\ (\ref{m_int}) and (\ref{quadrac}).
From Fig.\ \ref{REM:1} we can see that $m \to 0$ in this limit and so
Eq.\ (\ref{quadrac}) yields
\begin{equation}
y = \kappa - \sqrt{\kappa^2 - 1}
\end{equation}
where $y = \tanh \left (\gamma t_1 \right ) $ and $\kappa = J \sqrt{\ln 2}/\gamma$.
Since $y < 1$ for $\kappa > 1$,  $t_1$ is finite and then, taking the limit $t \to \infty$ in
Eq. \ (\ref{m_int}), we find
\begin{equation}\label{m_power}
m = \left ( - \frac{\ln 2 \ln y}{\kappa} \right ) ~\frac{1}{\gamma t} .
\end{equation}
This important result indicates that relaxation to  equilibrium, which is characterized by
the ground-state configuration together with a cloud of very close mutant configurations
(the quasispecies distribution) is given by a power law with exponent $-1$.

An estimate of the speed of evolution can be obtained by considering the prefactor
of $1/\gamma t$ in Eq.\ (\ref{m_power}). We find that this prefactor vanishes at the extremes
$\kappa = 1$ and $\kappa \to \infty$, and reaches a maximum at $\kappa \approx 1.810$
or $J/\gamma \approx 2.174$. (If we had included the curve for, say, $J/\gamma = 1.5$ in Fig.\ \ref{REM:1}
we could have observed this non-monotonic behavior there.) This parameter setting
corresponds then to the slowest convergence to equilibrium, i.e. the minimum speed of evolution.
The maximum speed is obtained by setting $\kappa \to \infty$ (or $J/\gamma \to \infty$) which
amounts to taking a vanishingly small mutation rate.

%------------------------------------------------------------------------
\begin{figure}
\centerline{\epsfig{width=0.52\textwidth,file=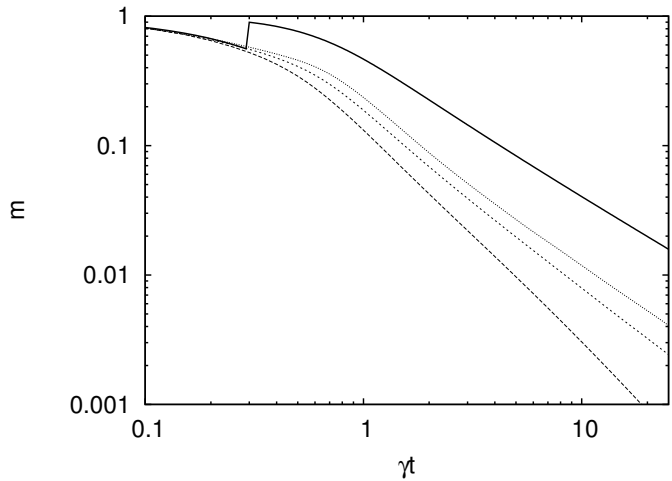}}
\par
\caption{Asymptotic dependence of the average overlap on $\gamma t$
for $J/\gamma=4$ and (broken lines from bottom to top) $N=6, 8$ and $10$.
The thick solid line is the theoretical prediction, which yields
$m \propto 1/ \left ( \gamma t \right )$  [see Eq.\ (\ref{m_power})]. }
\label{REM:7}
\end{figure}
%------------------------------------------------------------------------
%

To check whether a similar power-law scaling for large times
holds also for (infinite) populations of finite length sequences,
we solved the linear system (\ref{x_i}) through the direct diagonalization of $H$, which
is feasible only for relatively small sequence lengths. Figure \ref{REM:7}
summarizes our numerical results, which represent the average over $10^4$
independent samples. We find that the finite $N$ data is very well fitted
by the scaling law $m \propto \left ( \gamma t \right )^{-\alpha_N}$ with
$\alpha_6 = 1.77$, $\alpha_8 = 1.36$ and $\alpha_{10} = 1.17$. Assuming that
$\alpha_{\infty} = 1$ we find that these exponents,
in turn, are  described perfectly by  the function
$\alpha_N = 1 + 7.48 \exp \left ( - 0.38 N \right )$, which indicates
a very rapid approach to the value of the infinite-length exponent.

%%%%%%%%%%%%%%%%%%%%%%%%%%%%%%%%%%%%%%%%%%%%%%%%%%%%%%%%%%%%%%%%%%%%%%%%%%
%
\section{Ferromagnetic REM landscape}\label{sec:FREM}
%
%%%%%%%%%%%%%%%%%%%%%%%%%%%%%%%%%%%%%%%%%%%%%%%%%%%%%%%%%%%%%%%%%%%%%%%%%%

In the ferromagnetic REM \cite{de80,so89} we choose a particular
configuration, say $S^0 = \left (1, \ldots, 1 \right )$, and set its
fitness value to $J_0 N$. The other $2^N -1$ configurations are
assigned random fitness values $-E$ with $E$ distributed by the
Gaussian distribution (\ref{Gaussian}). From the evolutionary
modeling perspective, the new fitness level produces a gap in the
fitness landscape, which, as we show  here, results in nontrivial
dynamic consequences.

The quantum spin version of the ferromagnetic REM is defined by  the Hamiltonian
(\ref{Hamiltonian}) with the fitness function
\begin{eqnarray}\label{f_coup}
f \left ( s_1, \ldots, s_N \right ) &  =  &
\sum_{i_1 < i_2 \ldots < i_p} J_{i_1 \ldots i_p} s_{i_1} \ldots s_{i_p} \nonumber \\
&   & + N J_0 \left (\frac{1}{N} \sum_k s_k   \right )^p
\end{eqnarray}
where the multispin couplings $J_{i_1 \ldots i_p}$ are defined as in
Sect.\ \ref{sec:REM}. This fitness landscape is thus a linear combination of
the SP and REM landscapes. The equilibrium statistical mechanics of the quantum ferromagnetic
REM was studied  in Ref.\ \cite{sa93}, where the condition for the existence of
the ferromagnetic phase at zero-temperature (i.e., for $S^0$  be the ground-state configuration)
was found to be $J_0 > J \sqrt{\ln 2}$. Here we will consider only parameter settings that
satisfy this condition.

As in the previous cases, we use the decomposition of the transition amplitude
 $Z$, Eq.\ (\ref{dec}),  to solve the dynamics for the overlap $m$ between the initial
and the typical configurations at time $t$. The important change is that now the sum over the
final configurations $S^j$ in Eq.\ (\ref{dec}) does not include the master sequence $S^0$, which must be
considered separately. Hence we find that  $Z$ is given by a sum of two terms, the first is
the REM contribution,  Eq.\ (\ref{Z1}),  and the second is the SP contribution, Eq.\ (\ref{Z_0i}).
In particular, we focus on the case $m_0 = 0$ only, so that the latter equation becomes
\begin{equation}\label{Zf}
Z_F = \exp \left [ N \left ( \frac{1}{2} \ln \frac{\sinh \left ( 2\gamma t'_1 \right )}{2} -\gamma t
+ J_0 \left ( t - t'_{1} \right ) \right ) \right ]
\end{equation}
where $y' \equiv \tanh \left ( \gamma t'_1 \right )$ is given by the quadratic equation
$ \left ( y' \right ) ^2 - 2 J_0 y'/\gamma + 1 = 0$
(see Eq.\ (\ref{qua_SP}) with $m=m_0=0$), which has real solutions for $J_0 \geq \gamma$.
As the interesting situation is one where the spin-glass solution, given by Eqs.\
(\ref{m_int}) and (\ref{quadrac}), exists as well, so that $J/\gamma > 1/\sqrt{\ln 2}$, we
have
\begin{equation}\label{}
\frac{J_0}{\gamma}  >  \frac{J}{\gamma} \sqrt{\ln 2} > 1 ,
\end{equation}
so the existence of the ferromagnetic and spin-glass phases guarantees that $y'$ is real.

To obtain the time evolution of $m \left ( t \right )$ we first calculate
$\ln Z_{SG}$ using Eq.\ (\ref{Z_SG}) and $\ln Z_F$ using Eq.\ (\ref{Zf}) for fixed $t$.
(We take logarithms here because the relevant quantities are the expressions in the arguments of the
exponentials that define the transition amplitudes.)
If these two quantities are negative, we choose the diffusive solution, Eq.\ (\ref{m_D}).
If $\ln Z_F > \ln Z_{SG}$ then the system is in the ferromagnetic regime and so $m = 0$;
otherwise we choose $m$ given by the spin-glass solution, Eqs. (\ref{m_int}) and (\ref{quadrac}).

%------------------------------------------------------------------------
\begin{figure}
\centerline{\epsfig{width=0.52\textwidth,file=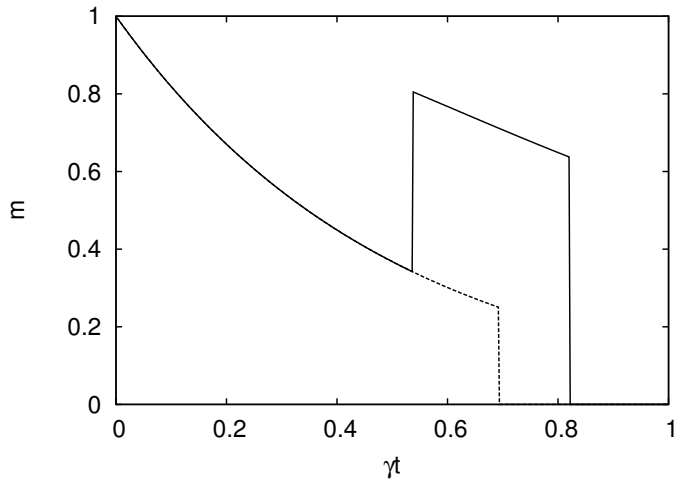}}
\par
\caption{Average overlap $m$ as function of scaled time $ \gamma t$.
The solid curve for $J_0/\gamma= J/\gamma = 3$ shows a  transition
between the diffusive and the spin-glass regimes at $\gamma t = 0.538$, and a
transition between the spin-glass and the ferromagnetic regimes at $\gamma t = 0.820$.
The broken curve for $J_0/\gamma=3$ and $J/\gamma = 2$  shows a situation where
there is a direct transition  between the diffusive and ferromagnetic regimes at $\gamma t = 0.693$. }
\label{FEM:1}
\end{figure}
%------------------------------------------------------------------------
%

\begin{figure}
\centerline{\epsfig{width=0.52\textwidth,file=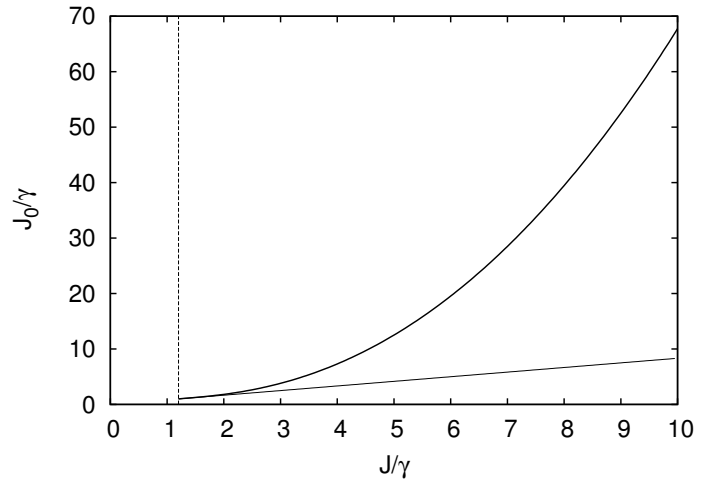}}
\par
\caption{The spin-glass regime interfaces the diffusive and the ferromagnetic regimes only
in the region of parameters located below the thick solid curve. This curve begins at the point
$ ( 1/\sqrt{\ln 2}, 1)$ and diverges as
$J^2$ for large $J$.  The thin solid straight line
is $J_0 = J \sqrt{\ln 2}$, below which the ferromagnetic phase is absent. }
\label{FEM:2}
\end{figure}

As the setting
$J_0 > J \sqrt{\ln 2}$ implies that the equilibrium phase is the ferromagnetic one,
one must have $m=0$ for large $\gamma t$. On the other hand, if the fitness of the initial
configuration is not greater than $\gamma$ (which we tacitly assume in this section), the
diffusive regime dominates for small $\gamma t$. The question is then whether an intermediate,
spin-glass regime appears between these extremes. The answer is given in Fig.\ \ref{FEM:1},
which indicates that the appearance of the intermediate regime depends on the values of the
parameters $J$ and $J_0$. To determine the region in the space of parameters
$\left ( J/\gamma, J_0/\gamma \right )$ where the spin-glass regime interfaces the other two
regimes, we have to calculate the value of $J_0/\gamma$ such that the time $t_{ds}$ at which
the transition from the diffusive to the spin-glass regime coincides with the time $t_{df}$
at which the diffusive regime transitions to the ferromagnetic one. Note that $t_{ds}$
depends only on $J$ as  shown in Fig. \ref{REM:3}. As for the time $t= t_{df}$ at which the transition
between the diffusive and the ferromagnetic
regimes takes place, it can be  calculated analytically by setting  $\ln Z_F = 0$ (see
\cite{sa04}). The final result is
\begin{equation}\label{tdf}
\gamma t_{df}= \frac{1}{4 \left ( \kappa_0- 1 \right )} \left [ \kappa_0  \ln \frac{\kappa_0+1}{\kappa_0-1} +
\ln \left [ 4 \left ( \kappa_0^2-1 \right ) \right ]
\right ]
\end{equation}
where $\kappa_0 = J_0/\gamma$. The procedure for searching  the values of $J_0/\gamma$,
for fixed $J/\gamma$, at which $t_{ds} = t_{df}$ is implemented numerically and the
result is shown in Fig.\ \ref{FEM:2}. Above the thick solid line there are only two dynamic
regimes, the diffusive and the ferromagnetic, and the time $t_{df}$ at which the transition occurs is
given   by Eq.\ (\ref{tdf}). In what follows we will concentrate on the study of the dynamics
for the parameters in the region below that line, where the three dynamic regimes are present.
In particular we will focus on the transition between the spin-glass and the ferromagnetic
regimes, which happens at time $t=t_{sf}$.

%------------------------------------------------------------------------
%------------------------------------------------------------------------
%

\begin{figure}
\centerline{\epsfig{width=0.52\textwidth,file=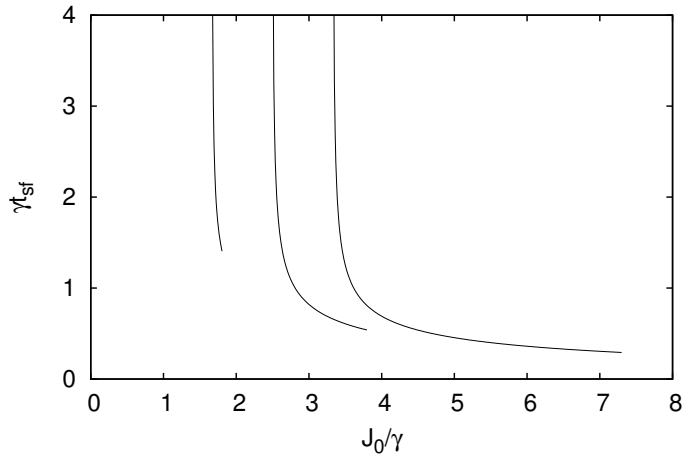}}
\par
\caption{Scaled time at which the spin-glass transitions to the ferromagnetic
for (left to  right) $J/\gamma = 2, 3$ and $4$. This transition occurs only within
a limited region of the parameter space, as illustrated in Fig.\ \ref{FEM:2}.
The divergences occur at $J_0 = J \sqrt{\ln 2}$.}
\label{FEM:3}
\end{figure}

Before we offer an analytical approximation to $t_{sf}$ it is instructive to
study numerically its  dependence on $J_0/\gamma$ and $J/\gamma$. This is shown in
Fig.\ \ref{FEM:2}, from where it becomes clear that $t_{sf}$ is defined
in a narrow region of the parameter space, determined by the conditions that the ferromagnetic phase exists and
that the spin-glass regime interfaces the diffusive and ferromagnetic regimes. Figure \ref{FEM:3}
shows the discontinuity of the overlap at $t_{sf}$. Since the overlap in the
ferromagnetic regime is zero, the jump $\Delta m$ is actually the overlap
in the spin-glass phase.

%------------------------------------------------------------------------
%------------------------------------------------------------------------
%

%------------------------------------------------------------------------
\begin{figure}
\centerline{\epsfig{width=0.52\textwidth,file=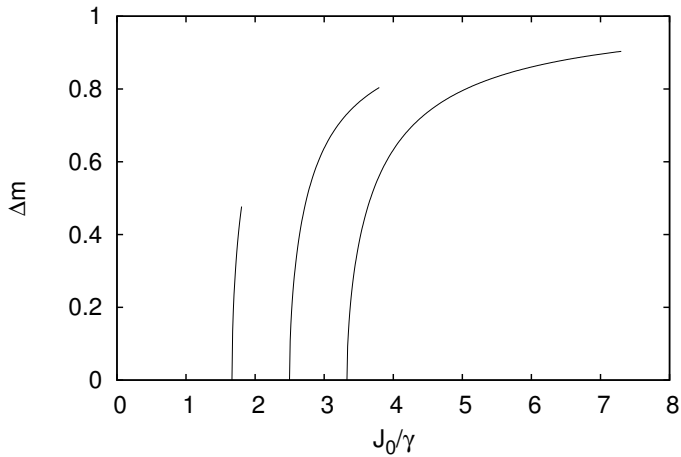}}
\par
\caption{The overlap discontinuity $\Delta m$ at $t= t_{sf}$ for (left to  right) $J/\gamma = 2, 3$ and $4$.
Since the overlap is zero in the ferromagnetic regime, $\Delta m$ equal the
overlap in the spin-glass regime.}
\label{FEM:4}
\end{figure}
%------------------------------------------------------------------------
%

The divergence of $t_{sf}$ and the vanishing of $\Delta m = m$ as $J_0$ approaches $J \sqrt{\ln 2}$ allows us to derive
an analytical expression for $t_{sf}$ in this limit. In fact, Eq.\ (\ref{m_power}) already provides an explicit expression for
$m$, namely,
\begin{equation}\label{ml}
m = -\frac{\sqrt{\ln 2} \ln y }{J t_{sf} }
\end{equation}
since $t_{sf}$ is large.
The equation that defines $t_{sf}$ is obtained by  equating $\ln Z_F / N$ to $F_{SG}$,
\begin{eqnarray}
\ln \left [ \frac{\sinh \left ( 2\gamma t'_1 \right )}{\sinh \left ( 2\gamma t_{1} \right )} \right ]
+ m \ln \tanh \left ( \gamma t_1 \right )  +  2 J \sqrt{h\left ( m \right )} t_1 & = & \nonumber  \\
+ 2 J_0 t'_1  -2 \left ( J_ 0 - J \sqrt{h\left ( m \right )} \right ) t_{sf} .
\end{eqnarray}
Recalling that for $\epsilon  \equiv \left ( J_0- J \sqrt{\ln 2} \right )/J_0 \to 0$ we have
$\kappa \to \kappa_0$ and so $t'_1 \to t_1$, we rewrite this expression as
\begin{equation}
t_{sf} = -m \frac{\ln \tanh \left ( \gamma t_1 \right )}{2 J_0 \epsilon}
\end{equation}
where we used $h\left ( m \right )\to \ln 2$ for $m \to 0$. Finally, inserting $m$ from
Eq.\ (\ref{ml}) into this expression yields
\begin{equation}\label{tsf_fin}
\gamma t_{sf} = - \left ( \frac{\ln 2}{2 \epsilon} \right )^{1/2}
\frac{1}{\kappa_0} \ln \left ( \kappa_0 - \sqrt{\kappa_0^2 - 1} \right ) .
\end{equation}
As $ m \sim 1/\left (\gamma t_{sf} \right )$ we find the typical mean-field
result $m \sim \epsilon^{1/2}$ at the transition.

Since $t_{sf}$ (or $t_{df}$, depending on the parameter settings) is the
waiting time for evolution to lead the system  close to its optimal fitness situation,
it is interesting to  see whether this waiting time can be minimized by a proper choice
of parameters (the mutation rate, for example). For fixed $\kappa_0$, we note that $t_{sf}$ must satisfy
the constraint $t_{sf} > t_{df}$ in the case the spin-glass regime is present. Hence the
minimum waiting time $t_{min}$ is obtained by equating the waiting times given in
Eqs. (\ref{tdf}) and (\ref{tsf_fin}). Of course, since $\epsilon \to 0$ we must set $\kappa_0 \to 1$
in the former equation. Keeping leading order terms in $\delta = \kappa_0 - 1$,
Eq.\ (\ref{tdf}) becomes $\gamma t_{df} \sim \left ( \ln 2 \right )/ \delta $, whereas
Eq.\ (\ref{tsf_fin}) reduces to $\gamma t_{sf} \sim \left ( \delta \ln 2 / \epsilon \right )^{1/2}$.
Equating these results yields $\delta = \left ( \epsilon \ln 2 \right )^{1/3}$ so that the minimum time
to reach the optimal fitness situation is
\begin{equation}\label{tmin}
\gamma t_{min} = \frac{ \left ( \ln 2 \right )^{2/3}}{\epsilon^{1/3}} .
\end{equation}
This expression is valid only in the limits $J_0/\gamma \to 1$ and
$J/\gamma \to 1/\sqrt{\ln 2}$.

%%%%%%%%%%%%%%%%%%%%%%%%%%%%%%%%%%%%%%%%%%%%%%%%%%%%%%%%%%%%%%%%%%%%%%%%%%
%
\section{ Conclusion}\label{sec:Conc}
%
%%%%%%%%%%%%%%%%%%%%%%%%%%%%%%%%%%%%%%%%%%%%%%%%%%%%%%%%%%%%%%%%%%%%%%%%%%

Most of the techniques from statistical
mechanics employed in the study of evolutionary models, such as Eigen's quasispecies model,
are manageable  only in the stationary regime $t \to \infty$ (see, e.g.,
\cite{Tarazona_92,fr93,Baake_97,Galluccio_97}).
Although the equilibrium analysis provides valuable insights into the
behavior of these models, a complete study of the dynamics is indispensable as
evolution is all about species dynamics, after all.

In this contribution we present an exact solution for the evolutionary
dynamics in an extremely rugged fitness landscape, Derrida's Random
Energy Model (REM) \cite{de80}. The evolutionary model studied is the quasispecies
model with a parallel mutation-selection scheme, in which mutations are
decoupled from replication \cite{Wiehe_95}. This scheme can be mapped in
Ising quantum chain in a transverse field \cite{Baake_97}, and the dynamics
can be solved using the Suzuki-Trotter formalism as done in the
case of the Single-Peak (SP) landscape \cite{sa04}.  In fact, the similarity
between the SP and REM-like fitness landscapes regarding their
steady-state distributions -- they are identical within the accuracy $\sim 1/\sqrt{N}$
-- is  well-known  \cite{fr97}, and here we explore it to  derive the evolutionary
dynamics on the REM landscape using the SP landscape results.

The (infinite) population is initially homogeneous, i.e., all sequences
are identical to a  reference sequence chosen such that its overlap with the
highest-fitness sequence is zero. In addition, most of our results are
based on the assumption that the fitness of this reference sequence is
negative. We note that in the parallel mutation-selection scheme,
we have a Malthusian fitness which basically measures the difference between
the reproduction and death rates, and so can take on positive and negative
values as well.

At each time $t$, the population is characterized  by the average overlap
with the reference sequence $m \left ( t \right )$ as well as by  the mean
fitness $R \left ( t \right )$. As expected, in the case the
initial configuration has low fitness (i.e., the fitness value is less than the mutation 
rate per site $\gamma$), the   dynamics for small $t$
corresponds to a random drift in the sequence space with $m$ decreasing exponentially
with increasing $t$ [see Eq.\ (\ref{m_D})]. Selection, which encodes information in the
fitness landscape,
has no role in the diffusive regime. We find, quite remarkably, that $m$ undergoes
a discontinuous transition at some finite $t = t_{ds}$ (see Fig.\ \ref{REM:1}) when the
dynamics enters a spin-glass regime in which $m$  vanishes as $1/t$ for large
$t$. As opposed to the SP landscape (and to the ferromagnetic version of REM; see below), the
dynamics needs an infinite time to reach the regions close to optimal fitness sequence.
When the initial configuration already has high fitness (i.e., the fitness value  is
greater than $\gamma$)
the diffusive regime is replaced by a pattern of stasis: the  dynamics  freezes
at the initial configuration (i.e., $m=1$ and $R = -E_i$) for a certain length of time $t_i > t_{ds}$
where  $ t_i = t_i \left ( E_i \right )$  
and then undergoes a discontinuous transition to the spin-glass regime.

In addition to the REM fitness landscape, we considered also the somewhat more
realistic  ferromagnetic-REM landscape which, as it is clear from Eq. (\ref{f_coup}),
can be viewed as a simple combination of the REM and SP fitness landscapes.
For some parameter settings (see Fig.\ \ref{FEM:1}), we find three distinct dynamic
regimes: the diffusive, spin-glass and the ferromagnetic regimes. The transitions
between these regimes are signaled by discontinuities of the overlap as well
as of the mean fitness. In a parameter setting such that the equilibrium phase
is the ferromagnetic one, the time to reach the optimal sequence is finite, as in the
SP case, but diverges near the (equilibrium) transition points. 
As in the case of
the ordinary REM, the diffusive regime is replaced by stasis when the initial configuration 
has a high fitness value.

The discontinuous  transitions between the different dynamic regimes are
similar to the punctuations, during which evolution proceeds very rapidly,  
observed in finite population  simulations \cite{Sibani_95,Aranson_97}. In fact,
one of the first theoretical models to reproduce the punctuated equilibrium
phenomenon made  explicit use of the effect of random genetic drift, which 
results from the finitude of the population, to promote the transition between
alternative fitness peaks \cite{Newman_85}. Since our results were derived
within the infinite-population assumption, this process cannot be responsible
for the observed punctuations. Alternatively, punctuations are
predicted by  models in which initially low frequency beneficial mutation 
becomes dominant in a few generations after a certain frequency threshold is
overcome \cite{Lensky_91,Johnson_95}. This is the process 
responsible for the punctuations observed in our model as well as in
microbial population experiments \cite{Elena_96}.

The infinite size population assumption  behind the quasispecies-like evolution
model considered here is a theoretical approximation only, and finite population size effects
are undoubtedly important. The discrete-time evolutionary dynamics on a
REM-like fitness landscape has been extensively investigated in the literature
for the finite population case \cite{kr03,kr05,de06,ja07,Wilke_02}.
Analytical approximations and numerical simulations have yielded many  interesting results about
record statistics and crossover
transitions. We note, however, that the exact solution of the deterministic model
exhibits a much richer dynamical structure. It
would be interesting to see whether there are any vestiges of the discontinuous 
transitions in the case
of finite but large populations and, in particular, how the coalescent time statistics
are affected by these regime changes \cite{Wilke_02}.

While our main interest is the investigation of the evolutionary dynamics,
our results bear on information theory as well \cite{ch85}, as they can be viewed as
the exact analytical solution for the decoding process (relaxation to the ferromagnetic
configuration)
of optimal codes \cite{so89,b4,b5,b6,b7}. In particular,
we conjecture that Eq.\ (\ref{tmin}) is universal for some classes of dynamics near
the error threshold-like transitions. In fact, the connection of evolution models with information
theory was first pointed out by
Eigen, who actually used  information theoretical arguments to derive an expression for the error
threshold in the SP landscapes \cite{ei71}. More recently, the relation between
molecular biology and  information theory was discussed in Refs.\ \cite{ad,Adami_04}.

In general, almost any fitness landscape can be qualitatively identified with one of three
classes: ferromagnetic, spin-glass and ferromagnetic spin-glass like.
The first class, which  exhibits a finite relaxation time to the optimum,
seems too simplistic to bear on real biological situations. The spin-glass
fitness landscape, on the other hand, exhibits an infinite relaxation time
to the optimum, which then could never be reached by the evolutionary dynamics.
The third class, which combines the complexity of the spin-glass landscape
with a finite relaxation time, seems to be the  preferable one from the
evolutionary perspective. Thus natural selection seems to choose the
type of fitness landscape that works more efficiently as
an information processing system.

The picture that emerges from computer experiments with digital
organisms \cite{ye02} resembles the case of SG fitness case.
Although this random macroevolution scenario may be described by a
spin-glass fitness landscape, Nature's preference seems to be for
the ferromagnetic spin-glass landscape, as manifested, for example,
by  protein evolution. In fact, it is known that proteins differ
substantially from random heteropolymers, and that random
heteropolymers can be described by the ordinary REM, whereas
biological polymers are described quite well by the ferromagnetic
REM \cite{wo87,gr00}. Hence the genome, which codes the information
to assemble the proteins, reveals ferromagnetic or ferromagnetic
spin-glass like fitness landscape. This is close to the idea of
channels in evolution \cite{de04}.  During the evolution, there are
large rearrangements of the genome, in addition to the point
substitutions considered here. This large transpositions resemble
the multi-scale optimization in computer science: perhaps Nature
takes advantage of large gene rearrangements, whenever the search
for new optimum by means of simple substitutions becomes too slow
\cite{b23}. These large events, as well as the simultaneous point
mutations in two or three adjacent sites, are permitted  because they
practically do not affect the error threshold, while the evolution
dynamics changes drastically, e.g., a relaxation time of about $10^6$
years in the case of single point mutations is reduced to 100 years
when triplet adjacent mutations are allowed \cite{b23}. We note that
simultaneous mutations in two or three random sites  yield the same
slow relaxation as in the case of a single point substitution.

In the theory of computation,  optimization problems are classified as polynomially solvable
if the  relaxation time to the optimum scales with some power
of the problem size,  and as NP-complete or as NP-hard  otherwise,
i.e., when the relaxation time scales exponentially with the problem size \cite{b22}.
Hence the channel-like evolution schemes are not
only ferromagnetic-type fitness, but also resemble the fast
computational schemes of polynomial class.

\begin{acknowledgments}
The work at Yerevan was supported in part by the VolkswagenStiftung grant ``Quantum Thermodynamics''.
The research at S\~ao Carlos was supported in part by CNPq and FAPESP, Project No. 04/06156-3.
D.B.S. thanks the
hospitality of the Instituto de F\'{\i}sica de S\~ao Carlos, Universidade de S\~ao Paulo, and the
FAPESP travel grant No. 08/10420-9 for  the support to his visit to S\~ao Carlos.
\end{acknowledgments}

%
%-----------------------------------------------------------------
%

%
%-------------------------------------------------------------------------------------------

\setcounter{equation}{0}
\appendix{}

\bigskip

%
%--------------------------------------------------
\section{Border contribution to  $Z$} \label{App_B}
%--------------------------------------------------
%
To complete the analysis of the integrals
in  Eqs.\ (\ref{Z_D}) and (\ref{Z_SG}) for large $N$, we must consider the contribution from  the border
$h \left ( m \right )= J^2 \left ( t-t_1 \right )^2/4$. Clearly, in this case we have
$Z_D = Z_{SG} = Z_B$ with
\begin{equation}\label{Z_B}
Z_{B} = \int _{-1}^1 dm ~ \exp \left [ N F_{B} \left ( m \right ) \right ]
\end{equation}
where
\begin{equation}\label{F3}
F_{B} = 2 h \left ( m \right ) + \phi \left ( m,t_1 \right) - \gamma t
\end{equation}
and $t_1 = t_1 \left ( m \right ) $ is a function of $m$ given by the border
equation. Maximization of $F_{B}$ with respect to $m$ yields the saddle-point
equation
\begin{eqnarray}\label{border}
\ln \left (\frac{1+m}{1-m} \right ) + \frac{1}{2}\ln \tanh \left ( \gamma t_1 \right ) & = & \nonumber \\
  \frac{\gamma\ln  \left ( \frac{1+m}{1-m} \right ) }{4 J \sqrt{h(m)}} \left [ \left ( 1+m \right )
\tanh \left ( \gamma t_1 \right ) +
 \frac{1-m}{\tanh \left ( \gamma t_1 \right ) } \right ]   &  &
\end{eqnarray}
which must be solved numerically. At the extremes $m = \pm 1$ we have $t_1 = t$ and so
$F_{B} = \ln \left [ 1 \pm \exp \left ( - 2 \gamma t \right ) \right ] - \ln 2 \leq 0$.
Our extensive numerical analysis  comparing the three exponents $F_{D} = 0$, $F_{SG}$ and $F_{B}$
indicates that whenever $F_B > 0$   we have $ F_{SG} > F_{B}$ and so the contribution from
the border can be neglected in comparison with those from the inner saddle points
discussed in the main text.

\end{document}